**Title: All van der Waals three-terminal SOT-MRAM realized by topological ferromagnet Fe$_3$GeTe$_2$**

*Jingyuan Cui[†], Kai-Xuan Zhang, [†]* and Je-Geun Park**

Jingyuan Cui, Dr. Kai-Xuan Zhang, Prof. Je-Geun Park
Center for Quantum Materials, Seoul National University, Seoul 08826, South Korea
E-mail: zaal@mail.ustc.edu.cn; jgpark10@snu.ac.kr

Jingyuan Cui, Dr. Kai-Xuan Zhang, Prof. Je-Geun Park
Department of Physics and Astronomy, and Institute of Applied Physics, Seoul National University, Seoul 08826, South Korea

Keywords: Topological layered ferromagnet Fe$_3$GeTe$_2$ (FGT), van der Waals magnetic tunnelling junction, three-terminal SOT-MRAM, and spintronics.

**Abstract:** Magnetic van der Waals (vdW) materials have attracted massive attention because of their academic interest and application potential for the past few years. Its main advantage is the intrinsic two-dimensionality, enabling much smaller devices of novel concepts. One particular exciting direction lies in the current-driven spin-orbit torque (SOT). Here, we, for the first time, realize an all vdW three-terminal SOT memory, employing the unique physics principle of gigantic intrinsic SOT of Fe$_3$GeTe$_2$ (FGT) and the well-known industry-adopted tunnelling magnetoresistance (TMR) effect. We designed the device operation procedure and fabricated the FGT/h-BN/FGT vdW heterostructure as a proof of concept. This device exhibits a classical TMR effect and unambiguously demonstrates the conception by precise performance as expected: the magnetic information of the top-FGT is written by current-driven SOT and read out by TMR separately. The writing and reading current paths are physically decoupled, enhancing the design and optimization flexibility substantially and further strengthening the device's endurance naturally. Our work would prompt more expansive use of vdW magnets for spintronic applications.

**1. Introduction**

Two-dimensional (2D) magnetic vdW materials[1-5] have attracted tremendous attention worldwide since their discovery and revolutionized the magnetism and 2D materials research fields in both fundamental science and device application. These vdW magnets provide incredible opportunities to scrutinize and explore intriguing fundamental physics such as the Mermin-Wagner theorem[4,6], monolayer Ising magnetism[2], BKT transition in monolayer XY magnetism[7], many-body entangled sharp magnetic exciton[8-10], to name only a few, at the true atomic two dimensions which was difficult to achieve previously. Meanwhile, by taking advantage of the 2D materials' superiorities[11,12], they also establish excellent platforms for functionalized device applications, including ferroelectricity-controlled magnetism of a ferroelectric/ferromagnet heterostructure[13], superconducting diodes with superconductor proximity to an insulating magnet[14-16], long-range skin Josephson supercurrent in a superconductor/ferromagnet/superconductor tunnelling junction[17], and particularly the spintronics— the most natural and practical application direction of a magnet toward next-generation highly-efficient chip devices.

Over the past few years, several types of devices have been demonstrated: giant spin-filter effect[18-20] in an insulating magnet junction, gate-tuned magnetism[21-25], gate-controlled exchange bias[26,27], large TMR in a metallic ferromagnet junction[28-31], and efficient spin-orbit torque[32,33]. FGT is particularly interesting in this aspect since it is the earliest discovered vdW ferromagnetic metal of perpendicular magnetic anisotropy[34] and hosts topological bands with large Berry curvatures[35]. As a result, FGT exhibits large anomalous Hall conductivity[35] and anomalous Nernst effect[36]. Another striking consequence of the topological Berry curvatures is the gigantic intrinsic SOT by a current in a single FGT itself[37-40], similar to the self-switching in $Fe_{2.5}Co_{2.5}GeTe_2$[41] system but in sharp contrast to the conventional SOT in an FGT/heavy-metal[32,33,42] or newly FGT/WTe$_2$[43-45] bilayer system. Benefiting from such a unique giant SOT[38,46], the self-switching by the current-driven SOT of FGT is highly energy-efficient and nonvolatile with multiple levels[39,47]. Despite the many spintronic demonstration cases of vdW magnets, the practical SOT- Magnetic Random Access Memory (MRAM) unit for a chip device remains unachievable, which otherwise would bring us closer to next-layer industrial applications.

In this work, we propose the idea of a three-terminal SOT-MRAM unit using all vdW heterostructure of top-FGT/h-BN/bottom-FGT. Current-driven gigantic intrinsic SOT can write the magnetic information in the top-FGT, which can then be read out by the industry-practical TMR effect of the whole tunnel junction, with the bottom-FGT serving as the magnetic pinning-layer. Following this device design principle, we manufactured the van der Waals magnetic tunnelling junction by the dry-transfer technique. The device shows the well-behaved TMR behavior with a TMR ratio of 6 %. Then, we realize the magnetization switching of the top-FGT free-layer by current-driven SOT using mA-order current and read out the information by TMR using a small current of μA-order after the writing process. The writing and reading processes are physically separated in this three-terminal SOT-MRAM, naturally enhancing the device's design and optimization flexibility and endurance. Our work conceptualizes and demonstrates the working model of three-terminal SOT-MRAM using all-vdW magnetic heterostructures and pushes forward the currently flourishing 2D magnetic materials toward the subsequent march of industrial application.

## 2. Results and Discussion
### 2.1. Physics principle and device operation procedure

As illustrated in **Figure 1a**, our previous works[38-40] discovered the gigantic intrinsic SOT by current in FGT itself without any heavy-metal layer directly related to its band topology and large Berry curvature. Such current-driven SOT is so unique that it can be readily incorporated into its free energy formula, modifying the spin-related free energy landscape. It can reduce the magnetic anisotropy and the resultant lowered energy barrier for the spin switching, eventually leading to a significant coercivity reduction by current. In other words, current-driven SOT in FGT can induce the hard-to-soft magnetic transition, resulting in the highly energy-efficient magnetic self-switching non-volatilely by current. Such intrinsic SOT in FGT has been sequentially reconfirmed using similar or different methods among independent research groups[46-50].

Figure 1b,c shows the proposed three-terminal SOT-MRAM unit, composed of the all vdW FGT/h-BN/FGT heterostructure, and its writing and reading functions by current. Similar to the typical magnetic tunnelling junction, the top-layer FGT represents the free-layer whose magnetization can be changed by external means, in our case, the writing current; the h-BN layer

serves as insulating spacer; the bottom-layer FGT works as the pinning-layer whose magnetization will not be altered. The writing and reading lines share the same ground line, forming a structure of three terminals in the device. The magnetic information of the free-layer top-FGT can be written by the current-driven gigantic intrinsic SOT highlighted in Figure 1a. On the other hand, the magnetization situations of the top-FGT can be read out by the well-known TMR effect illustrated in Figure 1d. Referred to the bottom-FGT pinning-layer, the tunnelling resistance depends on the relative spin direction of the top-FGT free-player and maximizes when the spins are aligned antiparallel for the top and bottom FGT.

The more detailed device demonstration procedures are described below. As depicted in **Figure 2**, we initialize the device by applying a magnetic field. Subsequently, the writing function can be achieved by applying an mA-order writing current to the top-FGT free layer with a small magnetic field below 100 Oe. Finally, we turn off the writing current and flow a small reading current of µA-order vertically to read out tunnelling magnetoresistance for the magnetic information.

The relative spin direction between the top-FGT free-layer and bottom-FGT pinning layer regulates the tunnelling resistance. It thus can reflect the magnetization information of the top-FGT free-layer. In this manner, the current path of the writing and reading process is physically separated in such a three-terminal SOT-MRAM. Such decoupling significantly enhances the design and optimization flexibility and, most importantly, strengthens the device's endurance since the reading current can be applied much smaller than the writing current.

**2.2. Device demonstration of TMR, writing and reading functions.**

As described in the Experimental Section, we fabricated the FGT/h-BN/FGT heterostructure device using the dry-transfer technique. **Figure 3a** shows the typical optical image of the fabricated device, highlighting the top-FGT, h-BN, and bottom-FGT layers, respectively. The thickness of top-FGT, bottom-FGT and h-BN was measured to be about 38, 18, and 2 nm, respectively (Fig. S1) by atomic force microscopy. As can be seen in Figure 3b, a classical TMR effect is observed in this heterostructure device: The tunnelling resistance gradually rises and develops a high resistance plateau when the magnetic field is swept from a positive to a negative direction (indicated by a solid black arrow and curve), which corresponds to the magnetic transition by a magnetic field from a parallel to an antiparallel spin configuration between the top-FGT and bottom-FGT layer. The magnetic transition and a subsequent resistance plateau also appear when the magnetic field is swept from a negative to a positive direction (highlighted by a dashed red arrow and curve). The TMR ratio, defined as (TMR-TMR(1T)) / TMR(1T), reaches a maximum of 6 %, which allows the following device demonstration of the writing and reading functions.

A passing remark: the magnetic moment of the free layer gradually changes during the flipping process, but the pinning layer shows a sharp transition in the device (Fig. 3b). Generally, whether the magnetic transition is sudden or gradual depends on the domain formation in the system. For FGT, multiple small domains dominate in thick FGT nanoflakes, while few big domains occur in thin FGT nanoflakes[51]. In our device, the free-layer FGT is around 38 nm thick, while the pinning-layer FGT is around 18 nm. Consequently, the free-layer's magnetization switches gradually, but the pinning-layer's spins flip sharply.

Several works in the literature focus exclusively on the TMR effect of FGT-based magnetic tunnelling junctions[28-31,52]. The TMR ratios can vary widely (e.g., between 0.2 % ~ 200 %) depending on many factors, such as insulating spacer material, spacer thickness, bias amplitude, and overall device quality[28-31,52]. In our measurements, the initial TMR ratio was approximately 33 % (Fig. S3), which is comparable to previous reports. However, it dropped to 6 % later after unavoidable electrical degradation, yet still exhibiting typical field-dependent TMR behavior. Despite being a smaller value, the 6% TMR ratio is sufficiently large for our central purpose to demonstrate the all-vdW three-terminal SOT-MRAM combining SOT writing and TMR reading functionalities.

**Figure 4a** exhibits the time sequence of the writing current and reading TMR values of all vdW three-terminal SOT-MRAM. After the writing current is applied, the measured reading TMR values remain step-like constants, revealing the switched magnetic states are nonvolatile. Extracted from Figure 4a, we plot the reading TMR as a function of writing currents in Figure 4b.

As can be seen, the reading TMR starts to rise above a critical switching current of 3.7 mA and boosts as the writing current increases, reflecting the gradual switching process by a current of the top-FGT's magnetization. Specifically, we indicate the switching process by current in Figure 4c in the reading TMR-magnetic field curve. The reading TMR climbs following the TMR-magnetic field curve under a tiny magnetic field of around -0.034 T (i.e., -34 Oe) when the writing current ramps are applied on the mA order. Such illustration visually demonstrates the writing function of the free-layer by the current-driven SOT non-volatilely, as well the reading function by TMR separately and reliably.

Unfortunately, we could not achieve complete magnetization switching even with a current of 8 mA. Notably, non-full switching by SOT has been frequently reported in numerous papers in the literature. Several factors can prevent full switching by current-generated SOT, which has been well summarized in a recent paper[53]: nonuniformity of magnetic domain nucleation[54], current-induced demagnetization due to Joule heating[32], geometry-induced domain-wall pinning[55], and interface contamination during the device fabrication[56]. On the other hand, FGT's intrinsic merit of high energy efficiency was maintained in our work: In this device for example, the top-FGT used as free-layer has a thickness of 38 nm and a width of 15 μm, so the current density corresponding to 8 mA is about $1.5 \times 10^6$ A/cm$^2$. Such a value is similar to the reported low value in FGT[38] and above two orders of magnitude smaller than the famous Pt/Co system[57,58].

This proof-of-principle work represents a prototypical all-vdW three-terminal SOT-MRAM, integrating the SOT writing and TMR reading in a complete device. However, we admit that there are still some limitations, such as the low working temperature and magnetic field required, which need to be improved further. For example, replacing FGT with other room-temperature 2D magnets[59,60] like $Fe_3GaTe_2$ will ultimately realize the room-temperature working device of the same geometry since $Fe_3GaTe_2$[49,50,59] was discovered to host an above-room-temperature Curie temperature while maintaining the same structural symmetries and physical properties. On the other hand, the small out-of-plane magnetic field in our device can be removed if an out-of-plane SOT or effective field is introduced to realize field-free switching, e.g., the unconventional out-of-plane spin Hall effect. There have been several attempts in spintronic fields to realize magnetic switching only by current-driven SOT with no magnetic field, even for the vdW magnet[45,53,61]. We hope to incorporate such field-free SOT switching in the future three-terminal SOT-MRAM device. Under that improved situation, magnetic field issue doesn't exist any longer in this device design.

## 3. Conclusion

We demonstrate a novel all-vdW three-terminal SOT-MRAM using topological vdW ferromagnetic metal FGT. Our device is based on the physics principle of gigantic intrinsic SOT in FGT and the well-known TMR effect. We also elaborate on the device demonstration procedure explicitly. Finally, we fabricated the FGT/h-BN/FGT vdW heterostructure and unambiguously work out such a conception with it. Our work brings 2D magnetic vdW materials, particularly FGT, to real applications and, most importantly, can inspire future attempts and endeavors to realize real industrial applications.

## 4. Experimental Section

*Sample preparation:* FGT single crystals were synthesized using the chemical vapor transport method[62,63], with iodine as the transport agent. The reactant was encapsulated in a vacuum tube and then placed into a two-zone furnace, with the temperature gradient set between 750 ℃ (source) and 680 ℃ (sink) for 7 days. h-BN single crystals were purchased from 2D semiconductors.

The FGT/h-BN/FGT device was fabricated using conventional exfoliation and polymer-based dry-transfer methods within an Argon-filled glove box. The top-FGT layer was initially exfoliated onto a Polydimethylsiloxane (PDMS) stamp and then directly placed onto an atomically thin h-BN nanoflake, which had been pre-exfoliated using scotch tape on a $SiO_2$ (90 nm)/Si substrate.

Subsequently, polycaprolactone (PCL) was used to pick up the FGT/h-BN heterostructure and drop it onto the bottom-FGT nanoflake pre-exfoliated on a SiO$_2$ (285 nm)/Si substrate. The PCL residue was then removed using Tetrahydrofuran (THF) solvent. No interface in this three-layer device made contact with any polymer during the fabrication of such a magnetic tunnelling junction.

The electrodes were patterned using standard electron beam lithography (EBL) with polymethyl methacrylate (PMMA A7). A Methyl isobutyl ketone (MIBK) / isopropyl alcohol (IPA) solution with a 1:3 ratio was used as the developer to remove the reacted PMMA. Argon plasma etching was applied to the developed device to eliminate the oxidized surface region of FGT and reduce the contact resistance. Subsequently, the sample was swiftly transferred to electron beam evaporation within 30 seconds, followed by 50/5 nm Au/Ti deposition.

*Electrical transport measurements:* The transport measurements were conducted on the Cryogen Free Measurement System (Cryogenic Limited) with a base temperature of 1.8 K and a magnetic field up to 9 T. Two sets of Keithley 6220 and Keithley 2182 were used for the writing and reading process of magnetic information, respectively.

**Supporting Information**
Supporting Information is available from the Wiley Online Library or the author.

*Note Added*
During the peer-review process, we appreciate the reviewer for bringing our attention to a related work, which was published very recently.[64]

**Acknowledgments**
J.C. and K.Z. contributed equally to this work. We thank Suhan Son and Inho Hwang for their support and helpful discussions. These works at CQM and SNU were supported by the Leading Researcher Program of the National Research Foundation of Korea (Grant No. 2020R1A3B2079375) and the Samsung Advanced Institute of Technology.

Received: ((will be filled in by the editorial staff))
Revised: ((will be filled in by the editorial staff))
Published online: ((will be filled in by the editorial staff))

**Figures**

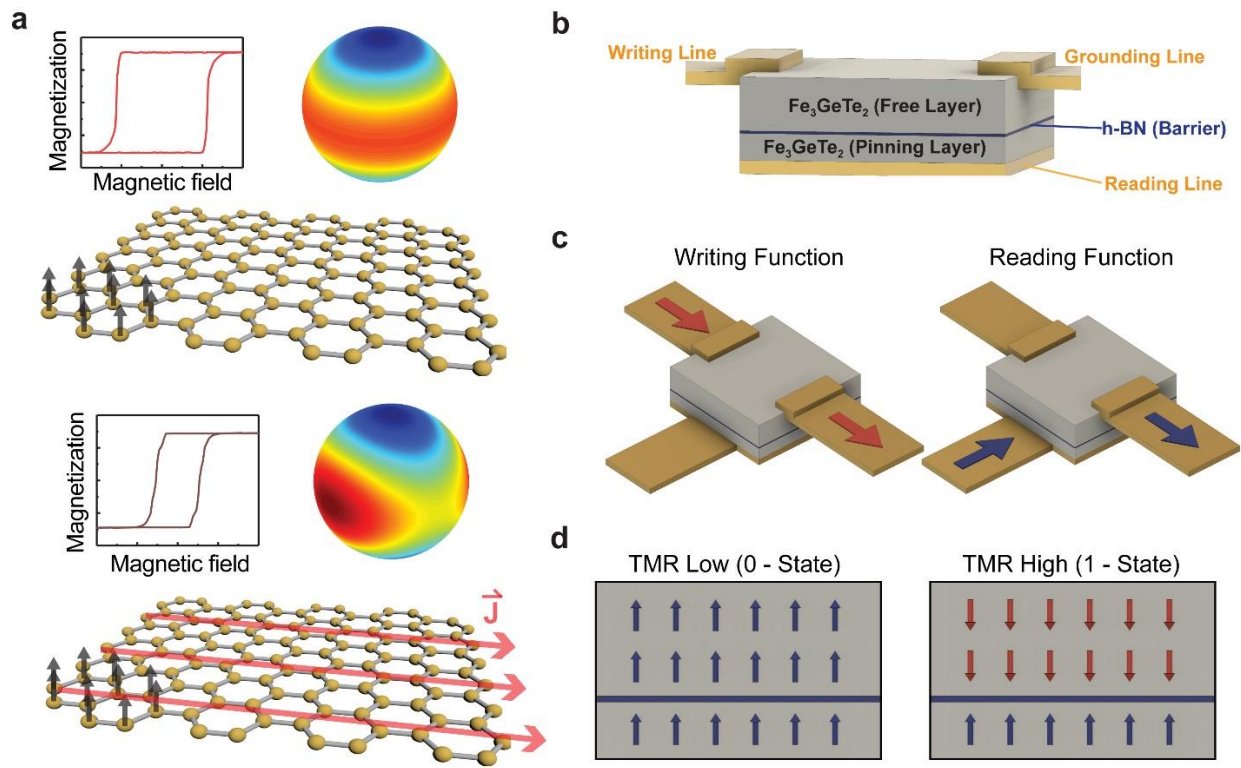

**Figure 1.** Physics principle of the proposed all vdW three-terminal SOT-MRAM. a) Current-driven gigantic intrinsic SOT in FGT itself can change the spin-related free energy landscape and reduce the magnetic anisotropy and, thus, the coercivity of magnetic switching substantially[38] (Some parts of the schematic are reproduced from our previous work[38]). b) The SOT-MRAM comprises top-FGT/h-BN/bottom-FGT heterostructures working as a free-layer/insulating-spacer/pinning-layer sandwiched tunnelling junction. The writing and reading lines share the same ground line, forming a structure of three terminals in the device. c) Separated current paths schematic for the writing and reading functions. Such three-terminal SOT-MRAM allows the physically decoupled current path, enhancing the design and optimization flexibility and endurance due to the much smaller reading current than the writing current. The current-driven

gigantic intrinsic SOT can write the magnetic information of the top-FGT in (a), which inherently exists in FGT itself without any heavy-metal layer. d) The top-layer FGT's magnetization can be read through the well-known TMR effect in a typical ferromagnet/insulator/ferromagnet tunnelling junction. The relative orientation between the top-FGT and bottom-FGT layers regulates the tunnelling resistance.

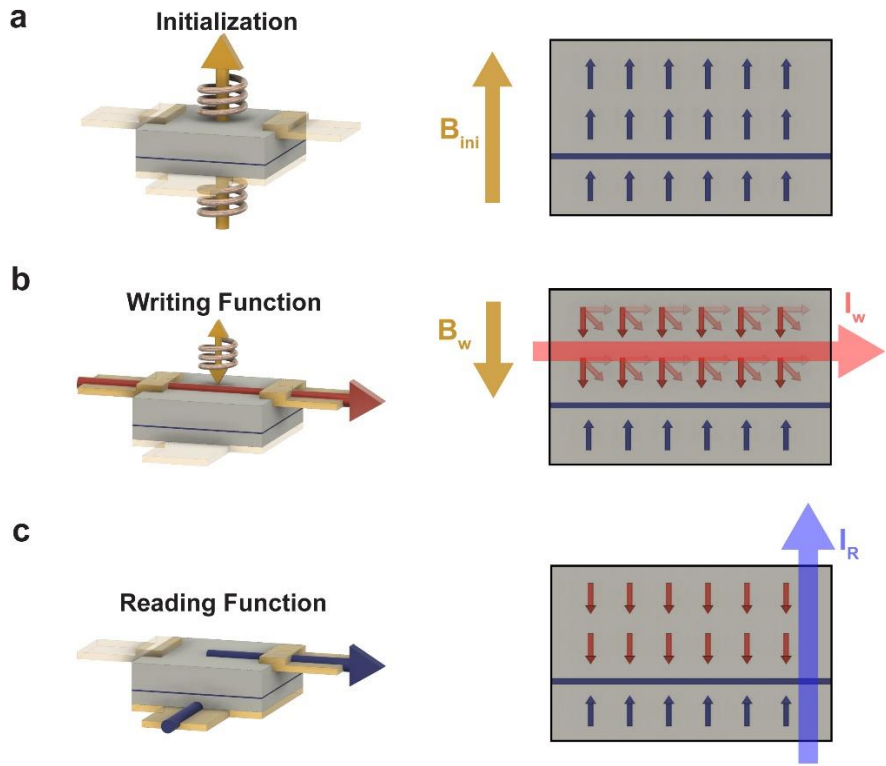

**Figure 2.** Device operation procedure. a) A magnetic field first initializes the device. b) For the writing function, mA-order writing current flows in the top-FGT layer under a small magnetic field below 100 Oe, where the current will flip the magnetization of the free-layer ferromagnet, i.e., the top-FGT layer by current-driven SOT. c) For the reading function, a small reading current of μA-order is applied vertically through the tunnelling junction. The measured magnetoresistance represents the magnetization information of the free-layer ferromagnet relative to the bottom-FGT pinning-layer.

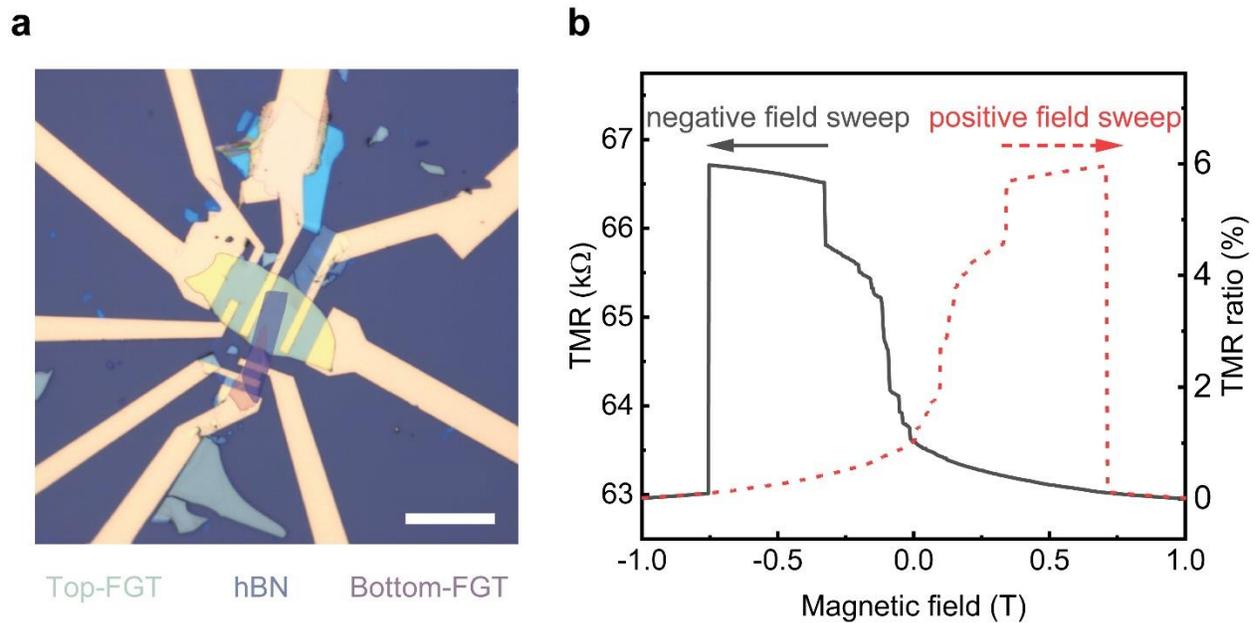

**Figure 3.** Device illustration and TMR effect. a) Optical image of the typical FGT/h-BN/FGT heterostructure device. The white scale bar is 20 μm. A few backup electrodes of more than three terminals were prepared in case of accidental breakdown of 2D material devices, which is necessary for high current tests. b) TMR effect of the demonstrating device at 2 K. High resistance plateaus develop while the magnetic field is swept from positive to negative (solid black curve and arrow) and from negative to positive (dashed red curve and arrow) directions. The TMR ratio is about 6 %.

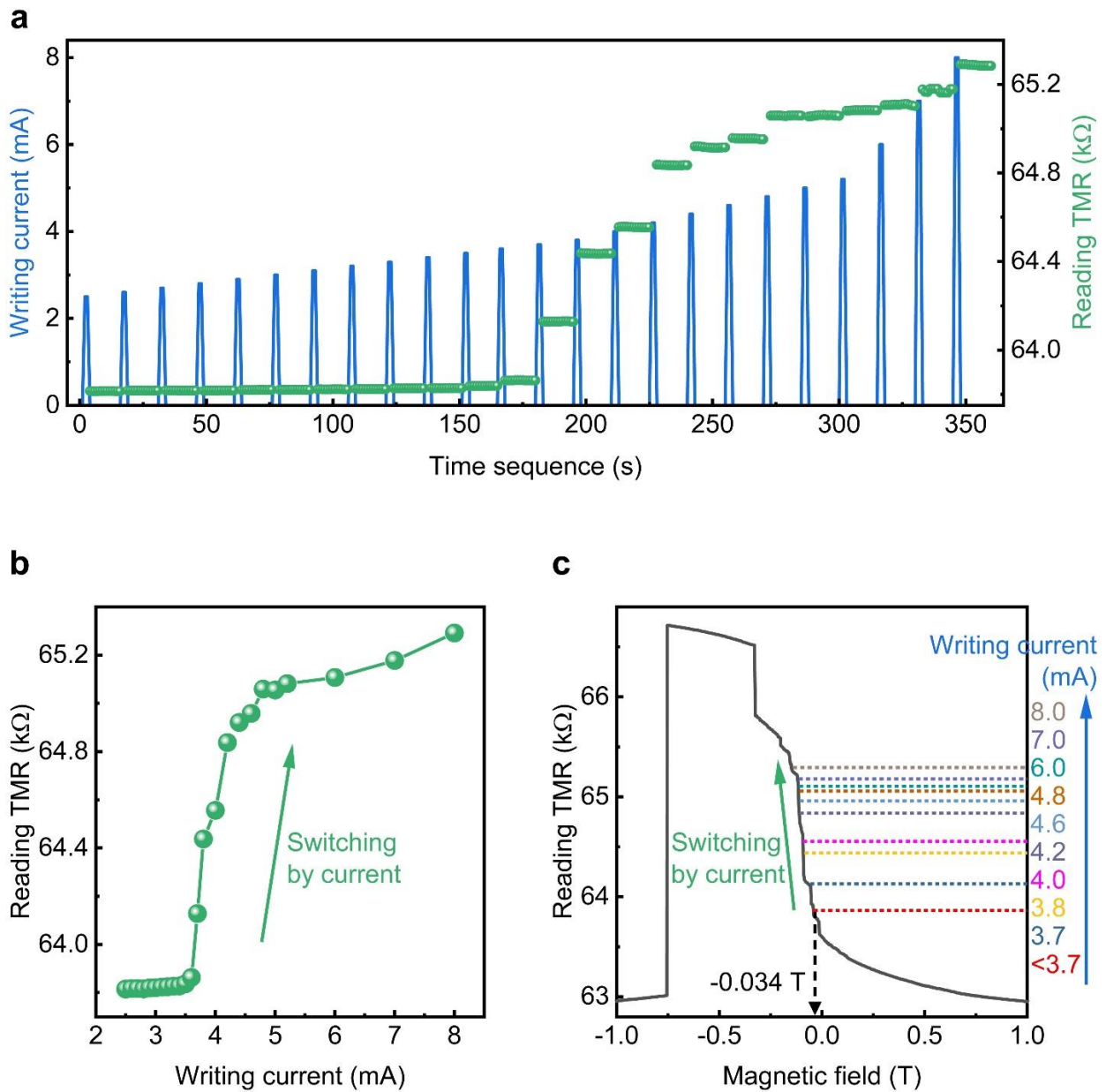

**Figure 4.** Function demonstrations of the all-vdW three-terminal SOT-MRAM unit at 2 K. a) Time sequence of the applied writing current and reading TMR. The writing current is on the mA order, and the reading current is 10 μA. The TMR value is constant after a writing pulse, featuring a nonvolatile magnetic switching. b) Reading TMR as a function of writing current extracted from (a). It rises significantly after the critical switching current of 3.7 mA and then shows an increasing trend as the writing current increases, representing the current-driven switching process. c) Writing and reading situations with the TMR reference. The TMR value climbs the reference TMR-magnetic field curve as the writing current increases under a small magnetic field of around -0.034 T. This presentation with the TMR reference demonstrates the achieved writing function by current-driven SOT and reading function by TMR separately.

WILEY-VCH

**The table of contents entry**

All van der Waals three-terminal SOT-MRAM is initiated using the FGT/h-BN/FGT heterostructure based on the gigantic intrinsic SOT of FGT and the tunnelling magnetoresistance effect. The device operation procedure is elaborated in sequence. The magnetic information of the top-FGT will be written by current and then read out by TMR separately. Such a concept is realized by the nanofabricated layered magnetic junction with well-behaved writing and reading performance. Our proposal and demonstration push the frontier of bridging novel 2D magnetic materials towards industrial applications.

Keywords: Topological layered ferromagnet $Fe_3GeTe_2$ (FGT), van der Waals magnetic tunnelling junction, three-terminal SOT-MRAM, and spintronics.

Jingyuan Cui, Kai-Xuan Zhang,* and Je-Geun Park*

Title: All van der Waals three-terminal SOT-MRAM realized by topological ferromagnet $Fe_3GeTe_2$

ToC figure

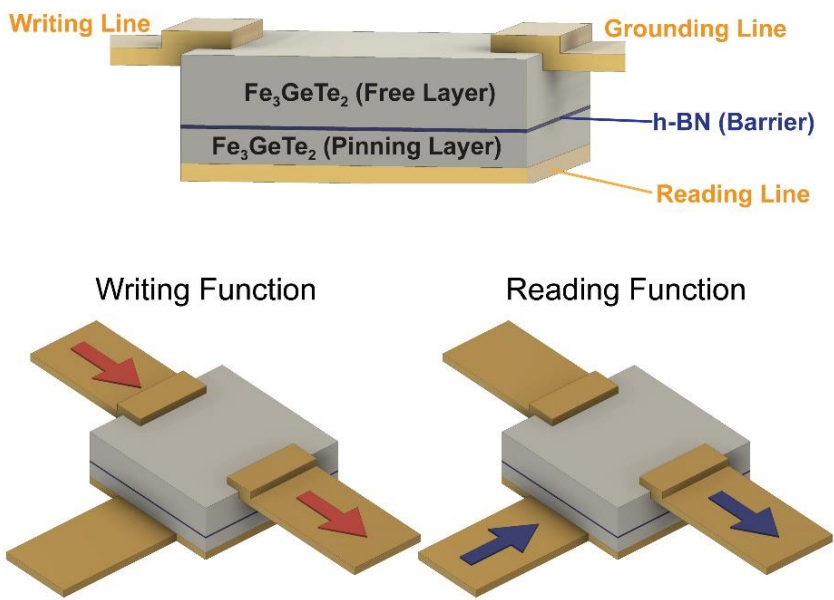

# Supporting Information

**Title: All van der Waals three-terminal SOT-MRAM realized by topological ferromagnet Fe₃GeTe₂**

*Jingyuan Cui, Kai-Xuan Zhang,\* and Je-Geun Park\**

Jingyuan Cui, Dr. Kai-Xuan Zhang, Prof. Je-Geun Park
Center for Quantum Materials, Seoul National University, Seoul 08826, South Korea
E-mail: kxzhang@snu.ac.kr; jgpark10@snu.ac.kr

Jingyuan Cui, Dr. Kai-Xuan Zhang, Prof. Je-Geun Park
Department of Physics and Astronomy, and Institute of Applied Physics, Seoul National University, Seoul 08826, South Korea

1. **Additional characterization of the Fe$_3$GeTe$_2$/h-BN/ Fe$_3$GeTe$_2$ (FGT/h-BN/FGT) device**

1.1 **Thickness characterization of the FGT/h-BN/FGT device**

The thickness of each layer was determined by room-temperature atomic force microscopy (AFM; Park systems, model NX10). The measurement was performed in contact mode after the transport measurements. The thickness of top FGT, bottom FGT and h-BN was measured to be about 38 nm, 18 nm, and 2 nm respectively.

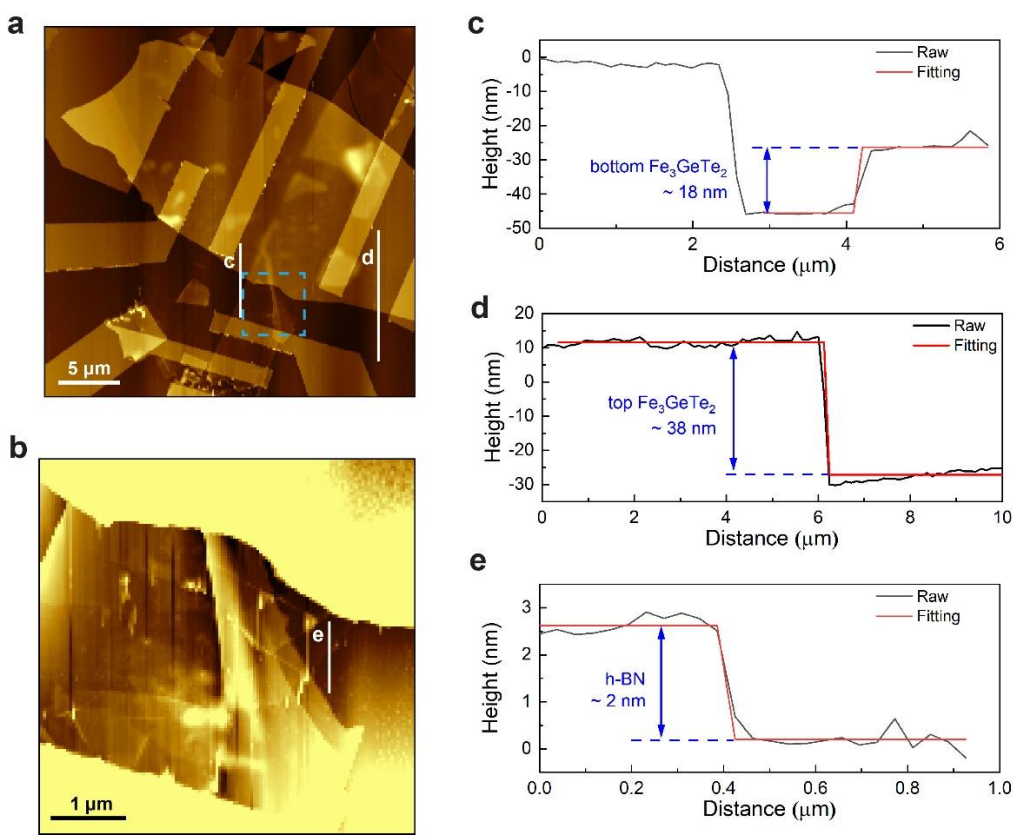

**Fig. S1.** Thickness information of the typical device in the main text. (a) The AFM image of FGT/h-BN/FGT device. (b) The AFM image of the area highlighted by blue square in (a). (c-e) The line profiles correspond to line c, d, and e in panels (a) and (b).

## 1.2 Bias dependent tunneling behavior of the FGT/h-BN/FGT device

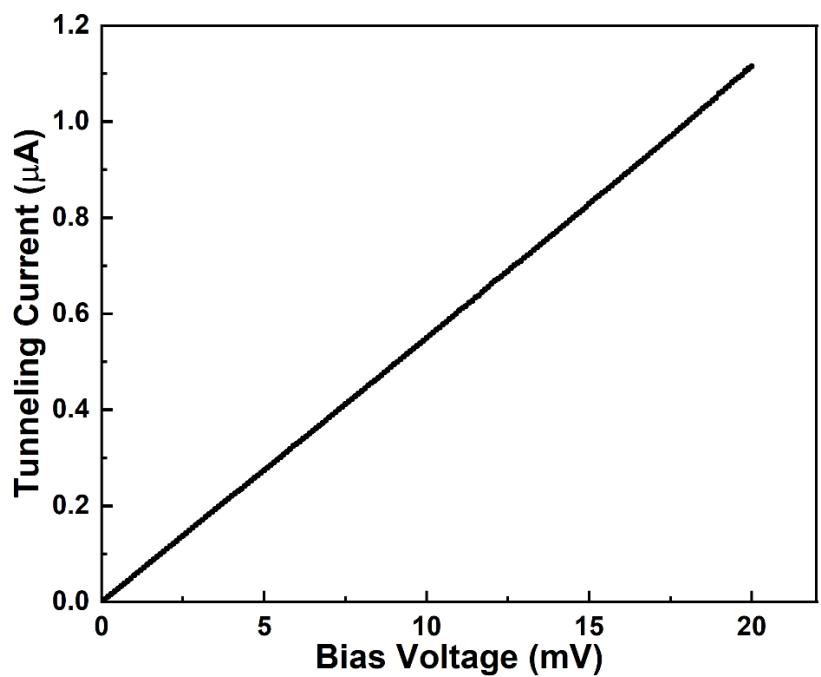

**Fig. S2.** Bias voltage dependent tunneling behavior measured at 2 K. The bias voltage was swept by the step of 0.1 mV. This nearly linear behavior has also been seen in many recent magnetic tunnelling junction works on layered magnet $Fe_3GeTe_2$ and $Fe_3GaTe_2$[1-3].

**1.3 Additional information of tunneling magnetoresistance (TMR) for the FGT/h-BN/FGT device**

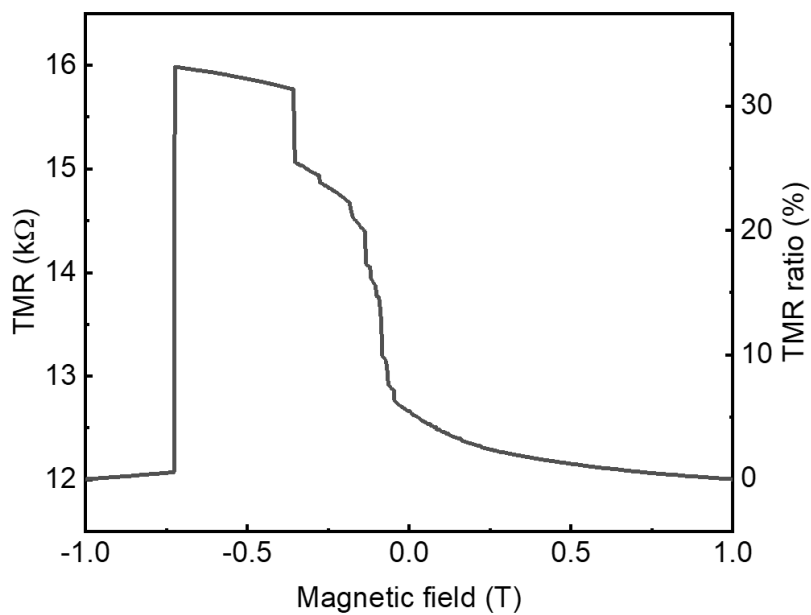

**Fig. S3.** The TMR of the FGT/h-BN/FGT device before certain electrical degradation. Such initial TMR ratio is about 33 %, within the reported TMR ratios ranging from 0.2 % to 200 % in literature [1-5].

**2. Possible exchange bias-like effect induced non-reciprocal behavior in the TMR curve.**

Note that the TMR effect presented in Fig. 3b is non-reciprocal: The "stage" in the negative field sweep curve locates at the field of -0.75 T, while the "stage" in the positive field sweep curve locates at the field being smaller than +0.75 T. We believe that the observed non-reciprocal behavior reflects the different coercivity of the bottom pinning-FGT layer during the magnetic field sweep, and can be attributed to an exchange bias-like effect. Exchange bias phenomena have been commonly observed in $Fe_3GeTe_2$-based heterostructures[6], gated nanoflakes[7], and also manifested in surface-oxidized nanoflakes[8] and in Te-rich $Fe_3GeTe_2$ nanoflakes[9].